\documentclass{article}
\usepackage{arxiv}
\usepackage[utf8]{inputenc} % allow utf-8 input
\usepackage[T1]{fontenc}    % use 8-bit T1 fonts
\usepackage{hyperref}       % hyperlinks
\usepackage{url}            % simple URL typesetting
\usepackage{booktabs}       % professional-quality tables
\usepackage{amsfonts}       % blackboard math symbols
\usepackage{nicefrac}       % compact symbols for 1/2, etc.
\usepackage{microtype}      % microtypography
\usepackage{lipsum}
\usepackage{graphicx}
\usepackage{multicol}
\usepackage{float}
\usepackage{enumitem}
\usepackage{siunitx}
\usepackage[backend=biber,style=numeric,sorting=none]{biblatex}
\setlist[itemize]{leftmargin=*}

\newcommand{\beginsupplement}{%
        \setcounter{table}{0}
        \renewcommand{\thetable}{S\arabic{table}}%
        \setcounter{figure}{0}
        \renewcommand{\thefigure}{S\arabic{figure}}%
        \setcounter{section}{0}
     }

\title{Addressing machine learning concept drift reveals declining vaccine sentiment during the COVID-19 pandemic}

\author{
 Martin M\"uller\\
  Digital Epidemiology Lab\\
  Ecole polytechnique fédérale de Lausanne (EPFL)\\
  1202 Geneva, Switzerland\\
  \texttt{martin.muller@epfl.ch} \\
    \And
 Marcel Salathé\\
  Digital Epidemiology Lab\\
  Ecole polytechnique fédérale de Lausanne (EPFL)\\
  1202 Geneva, Switzerland\\
  \texttt{marcel.salathe@epfl.ch} \\
}

\begin{document}
\maketitle

\begin{abstract}
  Social media analysis has become a common approach to assess public opinion on various topics, including those about health, in near real-time.
  The growing volume of social media posts has led to an increased usage of modern machine learning methods in natural language processing.
  While the rapid dynamics of social media can capture underlying trends quickly, it also poses a technical problem: algorithms trained on annotated data in the past may underperform when applied to contemporary data.
  This phenomenon, known as concept drift, can be particularly problematic when rapid shifts occur either in the topic of interest itself, or in the way the topic is discussed.
  Here, we explore the effect of machine learning concept drift by focussing on vaccine sentiments expressed on Twitter, a topic of central importance especially during the COVID-19 pandemic.
  We show that while vaccine sentiment has declined considerably during the COVID-19 pandemic in 2020, algorithms trained on pre-pandemic data would have largely missed this decline due to concept drift.
  Our results suggest that social media analysis systems must address concept drift in a continuous fashion in order to avoid the risk of systematic misclassification of data, which is particularly likely during a crisis when the underlying data can change suddenly and rapidly.
\end{abstract}

% keywords can be removed
\vspace{1.5cm}
% \newpage

\begin{multicols}{2}

\section{Introduction}
Supervised and semi-supervised Machine Learning algorithms are now ubiquitous in the analysis of social media data.
At the core of these algorithms is their ability to make sense of a vast amount of semi-structured real-time data streams, allowing them to automatically categorize or filter new data examples into, usually pre-defined, classes.
Multi-class text classification has been successfully used in public health surveillance, election monitoring, or vaccine stance prediction~\parencite{salathe2011assessing,bermingham2011using,brownstein2009digital}.
In recent years such algorithms have also been developed to mitigate the negative effects of social media, such as in the detection of cyber-bullying, hate speech, misinformation, and automated accounts (bots)~\parencite{reynolds2011using,davidson2017automated,shu2017fake,davis2016botornot}.

The microblogging service Twitter has played a central role in these efforts, as it serves as a public medium and provides easy access to real-time data through its public APIs, making it the primary focus of this work.
Twitter is well described as a classical example of a non-stationary system with frequently emerging and disappearing topical clusters~\parencite{costa2014concept}.
This poses problems for the aforementioned applications, as the underlying data distribution is different between training time and the time of the algorithm's application in the real world.
This phenomenon is known as concept drift~\parencite{schlimmer1986incremental} and can lead to a change in performance of the algorithm over time.

It is important to distinguish concept drift from other reasons for performance differences between training and testing, such as random noise due to sampling biases or differences in data preprocessing~\parencite{vzliobaite2010learning,webb2016characterizing}.
A classic example of concept drift is the change in the meaning of classes, which requires an update of the learned class decision boundaries in the classifier.
This is sometimes also referred to as real concept drift.
Often, however, an observed performance change is a consequence of a change in the underlying data distribution, leading to what is known as virtual drift~\parencite{widmer1996learning,tsymbal2004problem}.
Virtual drift can be overcome by supplemental learning, i.e.\ collecting training data from the new environment.
A good example are periodic seasonality effects, which may not be fully represented in the initial training data and only become fully visible over time.
However, in practice it is usually very difficult (if not impossible) to disentangle virtual from real concept drift, and as a consequence they are treated as the same effect~\parencite{vzliobaite2010learning}.

On Twitter concept drift might appear on very different time scales and at different rates.
Sudden shifts in a debate might be triggered by a quickly evolving news cycle or a catastrophic event.
Concept drift may also be a slow process in which the way a topic is discussed gradually changes over time.
A substantial amount of work has been dedicated to detecting and overcoming concept drift~\parencite{widmer1996learning,vzliobaite2010learning,elwell2011incremental}.
Three basic re-training procedures for overcoming concept drift have been proposed: (i) a time-window approach, (ii) an incremental model, and (iii) an ensemble model~\parencite{costa2014concept}.
In the time-window approach, a sliding window of recent training examples is used to train an algorithm.
In this approach, the algorithm ignores training data collected outside of that time window.
The incremental model, in contrast, uses all previously collected training examples to re-train the model.
Lastly, the ensemble model trains a model for each time window and uses the consensus of all previous models for future predictions.
As found in~\parencite{costa2014concept}, in the case of hashtag prediction on Twitter data, the incremental method gave the best results.

Although sophisticated methods have been proposed to estimate concept drift in an unsupervised way~\parencite{katakis2010tracking,yang2008conceptual}, in practice, a certain amount of re-annotation for both the detection and re-training of models seems unavoidable.
The decision about which of the newly collected data to annotate points to an exploration-exploitation dilemma, which is usually addressed in the context of an active learning framework~\parencite{settles2009active}.
The Crowdbreaks platform~\parencite{muller2019crowdbreaks} is an example of such a framework and has been built with the goal of exploring optimal solutions to this problem in order to overcome concept drift.

A change in the underlying data distribution might not necessarily have a negative impact on classifier performance.
It is conceivable, for example, that a polarisation in a debate on Twitter about a topic could even lead to an improvement in classifier performance.
It is therefore important to ask how much we should be worried about concept drift: even if model performance were to decrease, the real impacts on our analysis or interpretation might be negligible.

The consequences of concept drift are task-, environment-, and model-dependent~\parencite{vzliobaite2016overview}.
Here, we will address concept drift in the specific case of vaccine stance classification.
Vaccine stance classification on Twitter data has been widely studied and has shown promising links to vaccination decision making and vaccine uptake rates in different countries~\parencite{salathe2011assessing,bello2017detecting}.
The COVID-19 pandemic further emphasizes its importance, as evolving concerns about vaccines may significantly influence their effect~\parencite{johnson2020online,burki2020online}.

To the best of our knowledge, only one study directly addressed concept drift in vaccine stance classification.
In this study~\parencite{d2019monitoring} on tweets posted between September 2016 and January 2017 in Italian language, the authors did not find a substantial improvement of their model from incremental re-training before specific events.
Re-training was performed on 60 newly annotated tweets from seven manually selected events.
The authors conclude that either their original algorithm was already quite robust towards concept change, or that the newly collected training data was too small to see an effect.

Here, we use FastText~\parencite{joulin2016bag} and BERT (Bidirectional Encoder Representations from Transformers)~\parencite{devlin2018bert}, two commonly used models in social media text classification.
Most work on the topic of concept drift was conducted using classical machine learning models, to which also FastText belongs.
These types of models are very reliant on high-quality annotation data.
More recently, models of the transformer family, such as BERT~\parencite{devlin2018bert}, have been proposed, which require significantly less annotation data.
In what follows, we will examine whether these two models also share different concept drift characteristics.

The goal of this work is to emulate a typical social media analysis study, in which data is collected for a certain period of time, and a supervised machine learning model is trained on a subset of annotated data.
The model is then published and used to predict newly collected data.
First, we will try to answer whether or not concept drift can be observed, and if so, at what rate it occurs.
Second, we will investigate the influence of the study duration and the amount of annotation data used.
Lastly, we will examine to what extent concept drift influences the final analysis outcomes, in this case a sentiment index.

\section{Results}

\subsection{Observing concept drift}
\label{sec:observing_concept_drift}

Throughout the \num{1188} day observation period, starting on July 1st, 2017 and ending on October 1st, 2020, a total of \num{57.5}M English vaccination-related tweets were collected.
A random subset of \num{11893} tweets were annotated with respect to stance towards vaccines, which resulted in \num{5482} (46\%) positive, \num{4270} neutral (36\%), and \num{2141} negative (18\%) labels (for further details see methods section~\ref{sec:annotation_data}).
The dataset therefore bears clear label imbalance.

\begin{figure*}
\centering
\includegraphics[width=.8\textwidth]{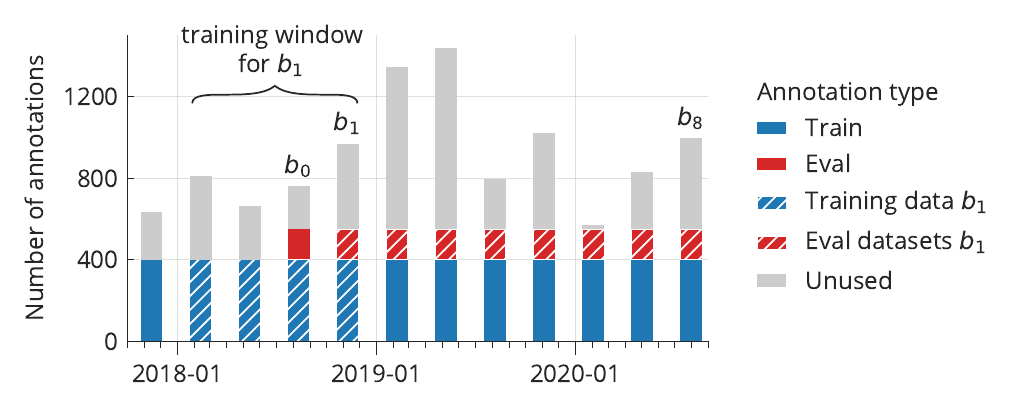}
\caption{
  Training and evaluation datasets.
  Each \num{90} day bin consists of \num{400} samples of training data (blue) and \num{150} samples of evaluation data (red).
  Each trained model is using the most recent \num{1600} samples for training, which is an equivalent of 4 bins or \num{360} days.
  For illustration purposes, the training data for the second bin $b_1$ is indicated as blue with white stripes.
  The $b_1$ model is then evaluated on all future evaluation datasets, indicated as red with white stripes.
}
\label{fig:1}
\end{figure*}

In order to observe whether classifiers experience drift in our dataset, we analysed the performance change of a model when predicting newly collected labelled data.
For this we used a sliding time window approach, as first proposed in~\parencite{costa2014concept}.
We dissected the collected \num{11893} annotations into 13 bins of 90 days each.
From each bin we sampled \num{550} examples and split them into a train ($n=400$, 72\%) and evaluation ($n=150$, 27\%) set (see Figure~\ref{fig:1}).
Each model was trained on a window of 4 bins of training data, which is equivalent to \num{1600} samples and a time span of \num{360} days.
The models are subsequently evaluated on the evaluation set corresponding to the bin at the end of their training window as well as on all future evaluation sets.
We repeat the process of binning, splitting, training and evaluating 50 times in order to yield a measure of confidence to our results.

\begin{figure*}
\centering
\includegraphics[width=.8\textwidth]{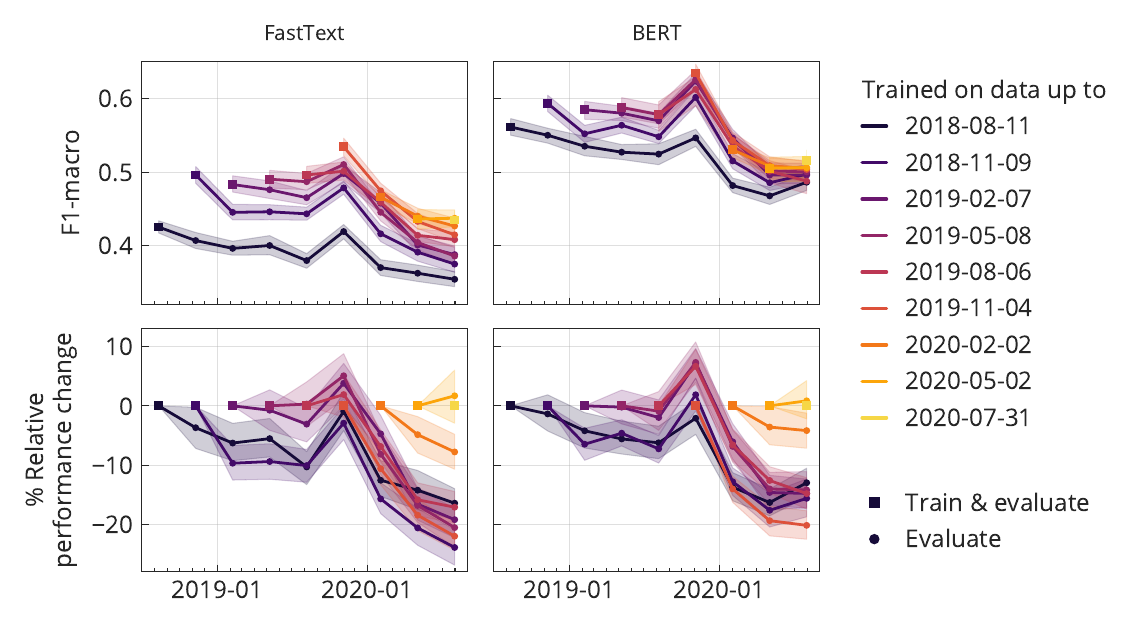}
\caption{
  Model performance over time.
  The top row shows absolute performance (in terms of F1-macro), and the bottom row shows the relative performance change of models compared to training time.
  The columns show the result for the two different classifier types FastText and BERT.
  The square indicates the performance at training time, the circles correspond to performance of that same model on future evaluation sets (compare with Figure~\ref{fig:1}).
  Bands correspond to bootstrapped \num{95}\% confidence intervals resulting from \num{50} repeats.
}
\label{fig:2}
\end{figure*}

Figure~\ref{fig:2} shows the classifier performance at training time (square symbol) and the performance at each future evaluation dataset (circle symbol) for classifiers trained on different training windows (color).
The upper left panel shows the results of these experiments for the FastText models.
We will first compare the initial performance in terms of F1-macro score (i.e.\ the arithmetic mean of the class-level F1 scores) of the classifiers on a test dataset which was sampled from the last bin of the corresponding training window (square symbols).
The initial performance of the first model is at \num{0.42}, the subsequent models plateau at around \num{0.50}, followed by a peak in fall 2019 with an abrupt decline in January 2020.
This variability in the initial performance of models points to considerable differences between training datasets over time.
The performance of the FastText models is quite low in general, which may be a consequence of the relatively small training dataset of \num{1600} examples and the lack of hyperparameter tuning.

Comparing the performance scores on future evaluation sets (circles) between models, we observe that the oldest models (black) generally perform worse than newer models (yellow) and that the ordering between models is preserved at all times.
However, in order to disentangle this effect from the variability in initial performance, we compute the relative change in performance with respect to performance at training time (lower left panel).
Starting from zero, the first model's performance drops quickly by around \SIrange{5}{10}{\percent}, followed by a rebound to initial performance in fall 2019, and ending in a sudden drop of approximately \num{-20}\% in early 2020.
The last drop indicates a very abrupt shift in concepts, twice as strong as during a comparable time window in 2019.
In fall 2019, changes in the data distribution allowed all models to rebound to initial performance, with some even ``over-performing'' by 5\% compared to training time.
This is a sign that the data distribution was particularly easy to predict.

Further investigation of the F1-scores by class reveals that concept drift is especially impactful on the negative class, whereas the positive and neutral classes do not experience a significant drift (see Figure~\ref{fig:S1}).
This could either indicate that anti-vaccine concepts are changing faster than pro-vaccine concepts or that the negative class is harder to learn due to label imbalance (cf.\ Figure~\ref{fig:S4}) and might, as a consequence, be more affected by virtual drift.
We will further investigate this difference in the next section.

Comparing these results to the BERT models (upper right panel), the models show higher absolute performance but they experience a similar level of relative performance loss and similar drift patterns.
This confirms that the observations are not model-specific but are likely to be observed in state-of-the-art semi-supervised machine learning models.

\begin{figure}
\centering
\includegraphics[width=\textwidth]{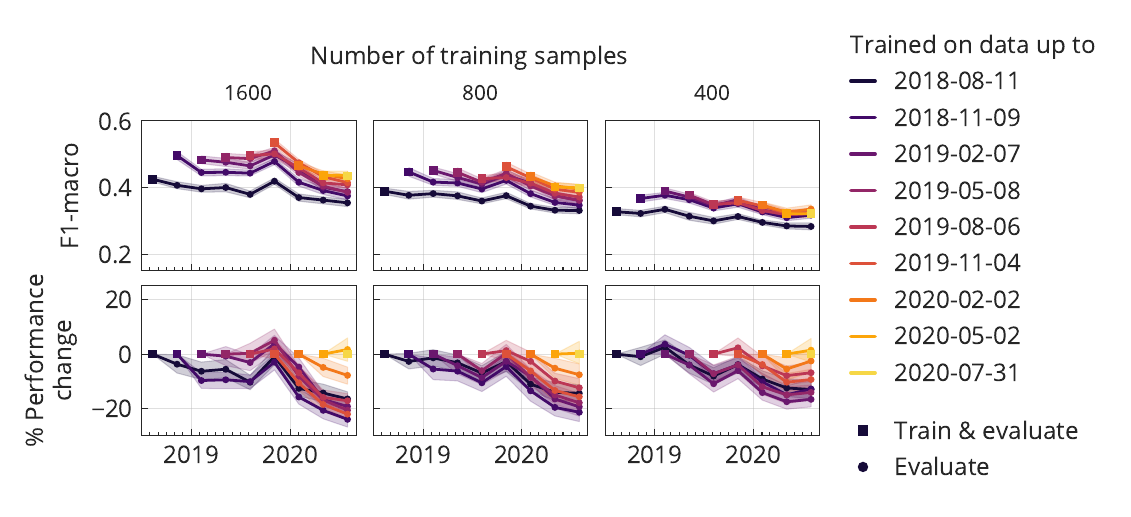}
\caption{
  Drift of FastText models depending on size of training data.
  The plots of the first column are identical to the FastText plots in Figure~\ref{fig:2}.
  For all experiments a training window length of 360 days was used.
  Initial performance is decreasing with a decreasing number of training samples.
  Overcoming concept drift is increasingly difficult, and is barely visible at 400 training samples.
}
\label{fig:S2}
\end{figure}

As previously stated, each model was trained on \num{1600} training examples over the previous 360 days.
Experiments were conducted under fewer training examples (Figure~\ref{fig:S2}) and smaller training windows (Figure~\ref{fig:S3}) for FastText.
As expected, training on fewer training examples leads to lower model performance, but we find the same drift patterns irrespective of the training data size.
Reducing the training window while keeping the number of training examples constant does seem to have an impact on performance or drift patterns.

\subsection{Explaining concept drift}
\label{sec:explaining_concept_drift}

\begin{figure*}
\centering
\includegraphics[width=.6\textwidth]{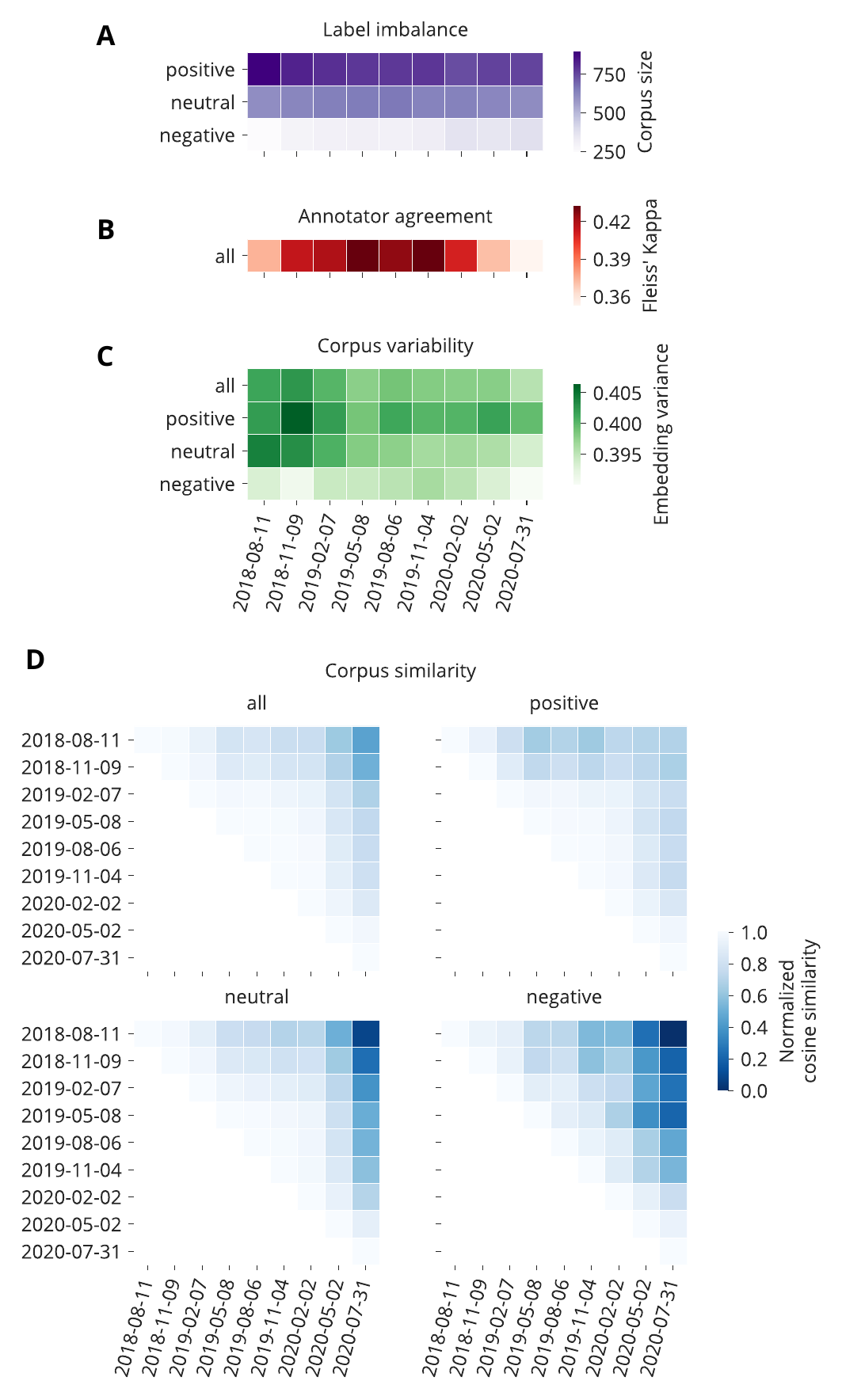}
\caption{
  Properties of the combined training data and first evaluation dataset for each trained model.
  \textbf{A.} Distribution in the number of labels per class.
  \textbf{B.} Annotator agreement, measured by Fleiss' Kappa.
  \textbf{C.} Corpus variability in terms of the variance of sentence embeddings within a corpus.
  Variability is shown for the full corpus as well by class.
  \textbf{D.} Normalized cosine similarity between the mean corpus vectors (i.e.\ the mean of all sentence vectors in each corpus) for all data as well as by class.
}
\label{fig:3}
\end{figure*}

\begin{figure}
\centering
\includegraphics[width=\textwidth]{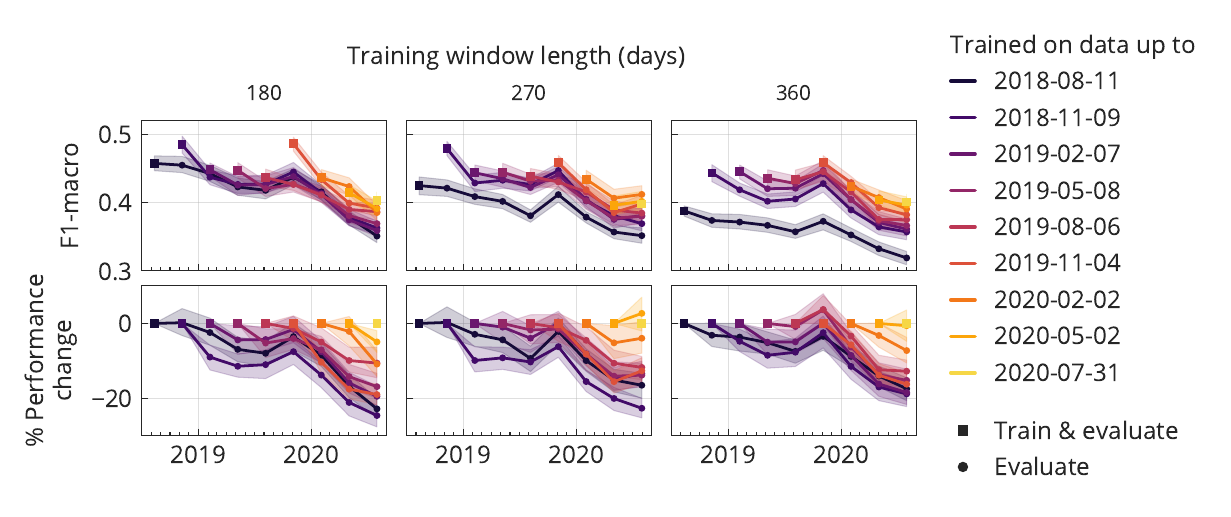}
\caption{
  Drift of FastText models depending on the length of the training window.
  Each model was trained on an equal number of 800 training examples, but distributed over 180, 270 or 360 days.
  A shorter training window is occasionally associated with slightly higher initial performance and slightly faster relative performance decrease on average.
}
\label{fig:S3}
\end{figure}

\begin{figure}
\centering
\includegraphics[width=.7\textwidth]{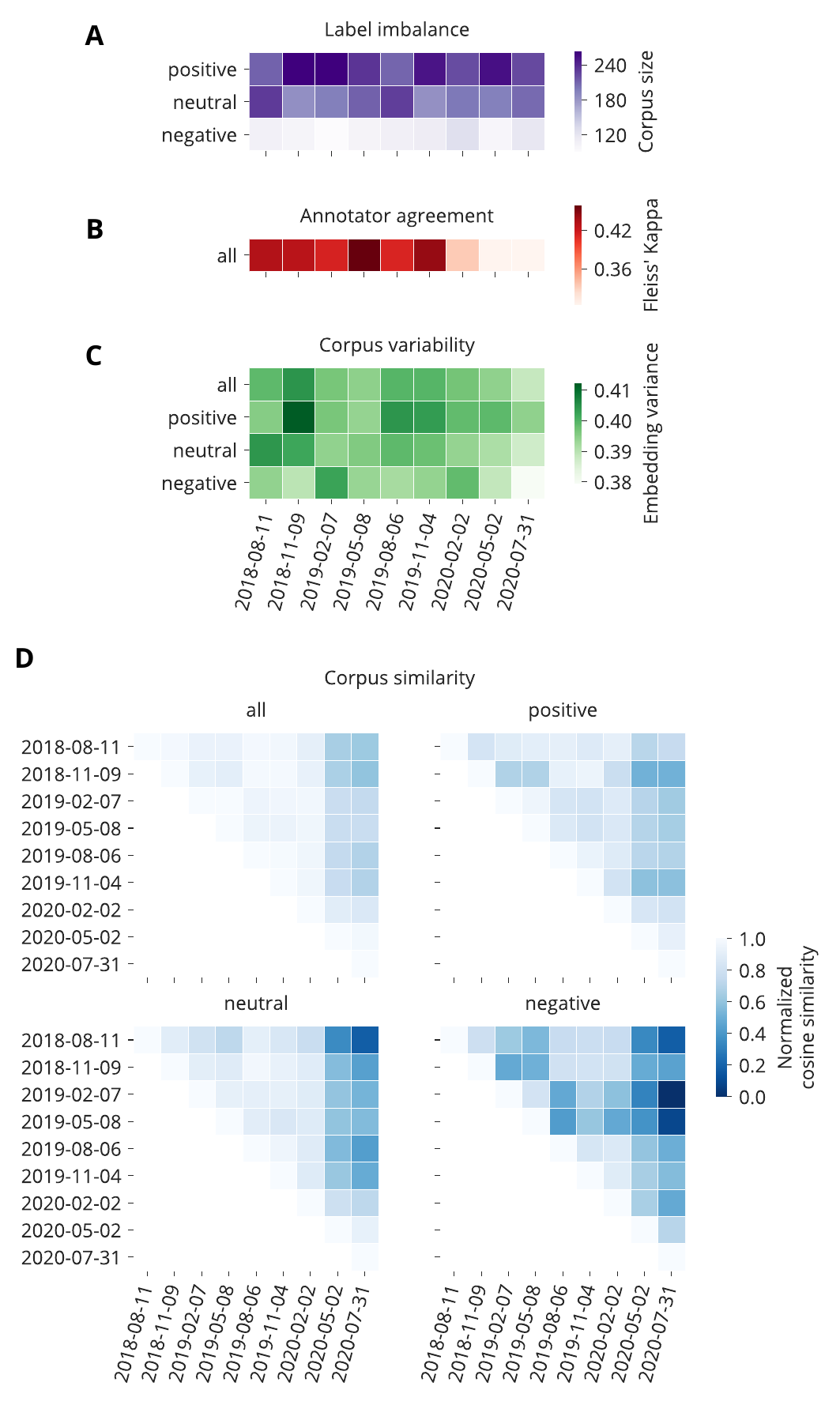}
\caption{
  This figure is equivalent to Figure~\ref{fig:3} in the main text, except for the different datasets that were used.
  In Figure~\ref{fig:3}, we show the used training and evaluation set in the full time window.
  This figure shows the newly added training and evaluation data for each 90 day bin.
  For a detailed description of this figure please refer to Figure~\ref{fig:3} in the main text.
}
\label{fig:S4}
\end{figure}

Next, we will try to explain both the variance in initial performance, as well as the different rates of drift observed.
We will investigate the effects of label imbalance, annotator agreement, and corpus variability on initial performance of models (Figure~\ref{fig:3}a-c).
Additionally, we compare corpus similarity over time and discuss it in the context of concept drift (Figure~\ref{fig:3}d).
In particular, we consider the first sampling (repeat) of the combined training ($n=\num{1600}$) and first evaluation set ($n=\num{150}$) for each training window.
The provided measures therefore correspond to what the model ``saw'' during training and in the first bin of evaluation. Figure~\ref{fig:S3} shows the equivalent metrics when limited to only the individual 90 day bins.

\subsubsection*{Label imbalance}
\begin{figure}[!h]
\centering
\includegraphics[width=\textwidth]{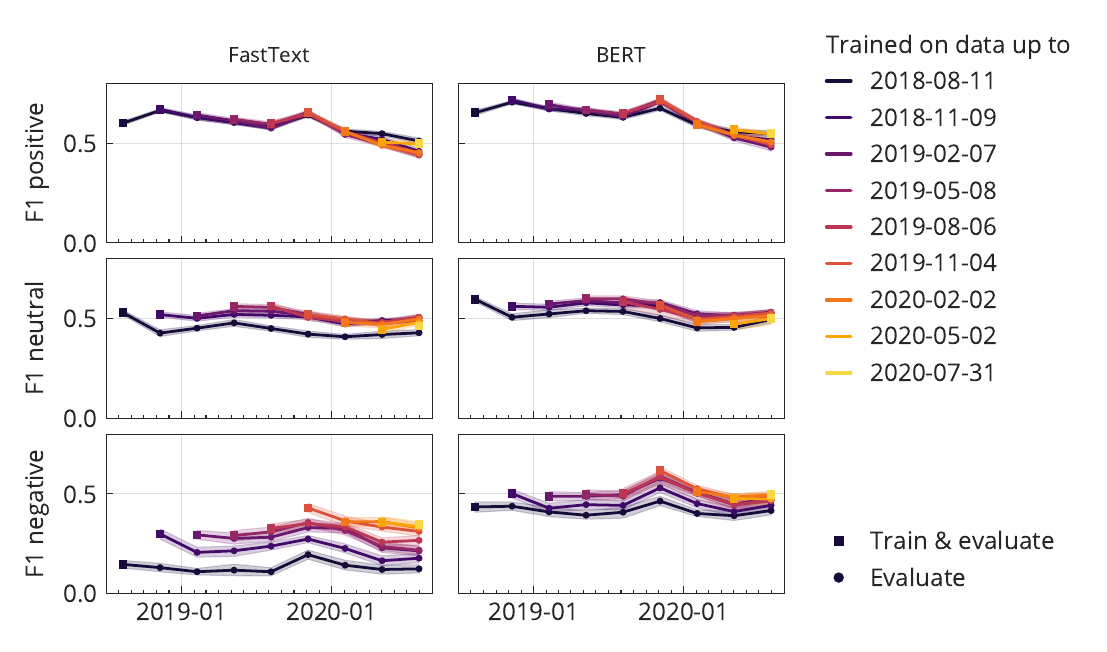}
\caption{
  Performance scores by class for FastText and BERT models.
  For an explanation of the Figure, please refer to Figure~\ref{fig:2} in the main text.
  Unlike for the negative class, performance between FastText and BERT is comparable for the neutral and positive class.
  The ``negative'' class shows the strongest effects due to concept drift.
}
\label{fig:S1}
\end{figure}

Although the used training datasets are always of equal absolute size, they vary in the number of examples per class over time (see Figure~\ref{fig:3}a).
It is commonly known that label imbalance can negatively impact model performance, which is also observed here (see Figure~\ref{fig:S1}).
However, we note that label imbalance was highest in the very beginning of the observation period and continuously decreased towards a more balanced situation.
Given the drop in initial performance in 2020, we conclude that label imbalance alone does not explain the observed variability in initial performance.

\subsubsection*{Annotator agreement}
We measure annotator agreement by computing the Fleiss' Kappa~\parencite{fleiss1971measuring} values for each dataset.
Annotator agreement is initially low at \num{0.37}, then increases to almost \num{0.45} and drops again to \num{0.36} in mid-2020.
This overlaps very well with the initial performance trend observed in Figure~\ref{fig:2}.
Variation in inter-annotator agreement may be a consequence of differences in annotation quality or difficulty of the annotation task, possibly hinting at semantic ambiguity of the text, as discussed next.

\subsubsection*{Corpus variability}
We use the BERT-large-uncased model to generate a 1024-dimensional sentence embedding vector (i.e.\ the vector of the CLS token) for each tweet text in the datasets.
Note that this BERT model has not been trained on any of our datasets, but it is able to generate rich sentence embeddings due to having been pre-trained on large amounts of English text.
Figure~\ref{fig:3}c shows the variance in the generated sentence embeddings across time.
We note that overall, corpus variability is highest in the beginning of our observation period, and then decreases towards the end.
Also, when considering the corpus variability by label class, we observe that negative samples have consistently lower variability compared to text labelled as positive.
The neutral class seems to undergo a shift from high to low variability.
In general, we may hypothesize that a lower variability points to lower separability in embeddings space, and therefore lower model performance.
This hypothesis aligns with the observations made in terms of initial performance.

\subsubsection*{Corpus similarity}
Similarity was measured by calculating the cosine similarity between the mean vectors for each corpus.
Low cosine similarity points to large semantic differences between datasets, which in turn could be an indicator for concept drift.
In the top left panel (``all''), the datasets are compared with each other.
We observe that over time, corpus vectors are moving further away from each other.
The biggest difference was observed between the two datasets furthest from each other in time (2018-08-11 and 2020-07-31).
We also observe a bright area in the middle of the heatmap, which reveals that datasets between February 2019 and February 2020 are more similar to each other compared to datasets before (2018) or after (May \& July 2020).
This aligns well with the results in Figure~\ref{fig:2}: Most of the concept drift was observed in 2018 and following 2020, whereas models in 2019 didn't drift by a lot.
When considering the corpus similarity by class, we can attribute most of these effects to the neutral and negative class.
We therefore show that anti-vaccine content ``drifts'' faster than pro-vaccine content.

In conclusion, our observations point to the fact that the differences in initial performance of models are likely a consequence of low annotator agreement.
The reason for this low agreement could be rooted in semantic ambiguity, as expressed by annotator agreement and corpus variability.
The degree of concept drift on the other hand is best explained by our measure of corpus similarity.

\subsection{Consequences of concept drift on real-time monitoring}
\label{sec:concept_drift_consequences}

\begin{figure*}
\centering
\includegraphics[width=.7\textwidth]{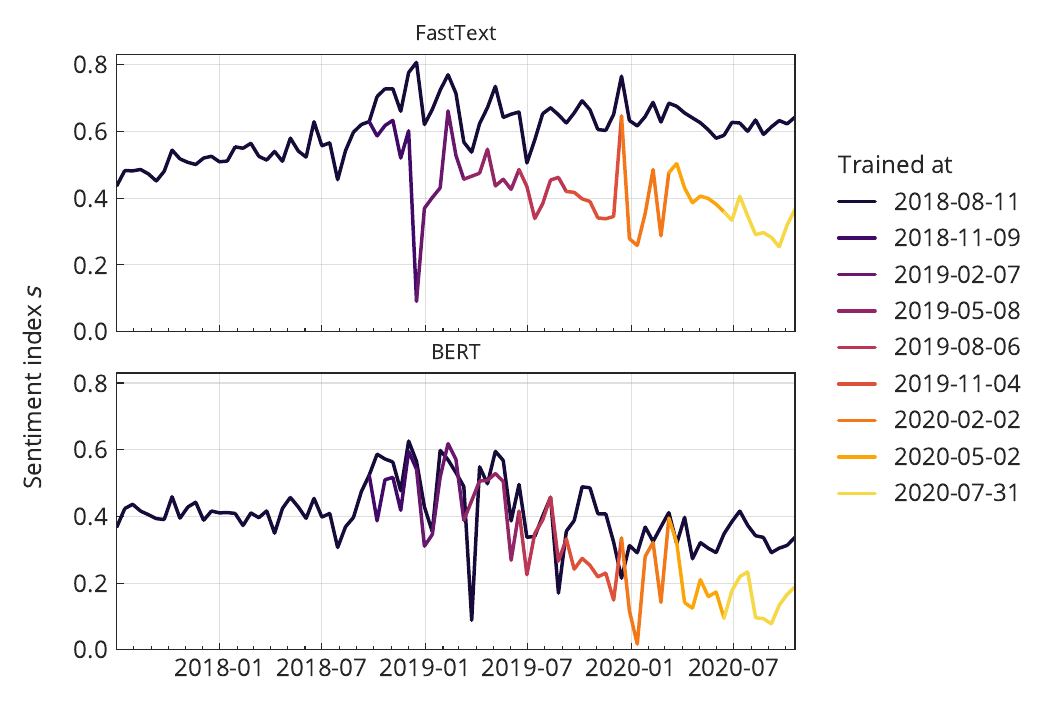}
\caption{
  Impact of concept drift on the predictions made by FastText and BERT models.
  Each panel shows the comparison of a model which was trained in August 2018 (black) to a model which was continuously updated every 90 days (colored).
}
\label{fig:4}
\end{figure*}

Lastly, but perhaps most importantly, we highlight the impact of concept drift on the inference of the previously trained models when used for real-time monitoring of new data.
We compare the predictions of a legacy model, which was trained in August 2018 and used for the two subsequent years, to a model we update (re-train) every 90 days.
We compute the sentiment index $s$, which corresponds to the weekly mean of positive, neutral and negative predictions, when mapped to the numerical values of \num{+1}, \num{0} and \num{-1}, respectively.
Figure~\ref{fig:4} shows these sentiment trends for both the FastText and BERT model variants.
We observe that, in the case of FastText, the sentiment predicted by the legacy model increased slightly until 2019 and then remained static.
The updated models, however, show a downwards trend starting in mid-2019 and dropping further in 2020.
By the end of our observation period the legacy model predicts a 0.3 points higher sentiment than the up-to-date models, while completely missing out on the downwards trend.

BERT models show a similar but smaller error, which is in agreement with our previous analysis.
The legacy BERT model was in agreement with the updated models at the time of the first drop in 2019, but then started to diverge.
We can therefore conclude that due to their higher overall performance, BERT models will have less severe deviations, but are not immune to effects of concept drift in the long run.
We also note a large difference in the extent of positive and negative spikes between the legacy and re-trained models.
Drift may therefore not only affect the mean sentiment trend but also sensitivity on shorter time scales, which could be problematic for real-time event or anomaly detection.

As previously stated, the sentiment trend of both the updated BERT and FastText models show a negative trend of the vaccine sentiment.
Given the current debate surrounding novel vaccines for the Sars-CoV-2 virus, this finding is concerning from an epidemiological perspective.
Note however, that the BERT models used for these predictions are of mediocre performance and future studies will be needed to confirm and interpret these trends.

\section{Discussion}

In this work, we investigated the effects of concept drift in vaccination-related Twitter data streams over a duration of three years.
Using a sliding time window approach, we emulate a social media study in which (i) data is collected for one year, (ii) an algorithm is trained, and (iii) the algorithm is used in real-time monitoring of new data.
While this may correspond to a common setup in social media analytics, we demonstrate here that without taking concept drift into account, the quality of the results will decay.
Using a vaccine-related dataset from 2018--2020, we demonstrate how failing to take concept drift into account would have largely missed a rather dramatic decay in vaccine sentiment during the COVID-19 pandemic in 2020.

We find that overall, concept drift indeed occurred, which led to a decline in model performance of over 20\% in the course of three years.
However, most of this decline happened in only ten months.
Concept drift therefore affected model performance at different rates throughout the observation period.
Furthermore, the relative performance loss was not consistently negative but reverted to initial levels, or even slightly above that.
These findings are consistent with the various ways real and virtual concept drift can occur.
Although BERT models yielded higher performance scores, they are not immune to issues related to concept drift.
On a relative scale, BERT models show the same degree of drift as the much less sophisticated FastText models.

In order to better understand the reasons for these phenomena, we investigate the properties of the used datasets.
We can explain the large differences in initial performance of models with differences in semantic ambiguity of the text, as indicated by low inter-annotator agreement and low corpus variability.
Occurrence of concept drift could be linked to differences in corpus similarity.
In particular, we find that the negative class is responsible for most of the decay in performance over time and also shows the strongest signs of drift.
Anti-vaccine content may therefore change topics at an increased rate compared to both positive or neutral content.

A caveat of this study is that the results are based on classifiers of mediocre performance.
Given the fact that the negative class was most affected by concept drift and is at the same time also the smallest class in our dataset, it is a fair question to ask whether concept drift would disappear given more annotation data and higher performance of models.
It is conceivable that more annotation data would lead to a better representation of the training window.
However, as results in a study on automated geo-location of tweets show~\parencite{dredze2016geolocation}, concept drift will still occur also under vast amounts of annotated data and adaptive re-training on even a relatively small corpus can overcome this drift.

Our results do not overlap with a previous study on vaccination-related Twitter data~\parencite{d2019monitoring}, which did not find concept drift in an observation period between September 2016 and January 2017 in Italian language.
The reason for this could be that the time scale analysed was too small to see an effect, or that concept drift was much smaller in that particular dataset.

It is safe to assume that the COVID-19 pandemic led to severe topical shifts in the vaccine debate, which ultimately translated into strong concept drift and model performance loss.
Based on these results, it can be expected that future crisis situations would lead to similarly strong concept drift, thereby severely undermining the utility of social media monitoring tools that do not take concept drift into account.
This is especially true for applications which are intended to be used exactly in such circumstances.

Although our work focused on the singular task of vaccine stance prediction, we believe that these results stress the general importance of addressing concept drift in any real-time social media monitoring project.
Overcoming concept drift is a complex task, and many algorithmic solutions have been proposed.
However, in order to succeed in practice, a tightly coordinated and fine-tuned framework for both the annotation and retraining of models is required.
The Crowdbreaks platform~\parencite{muller2019crowdbreaks} was built with the intention to address this issue and provide solutions for it.

\section{Materials and methods}
\label{sec:methods}

\subsection{Data collection}
\label{sec:data_collection}
This study is based on Twitter data collected through the Crowdbreaks platform~\parencite{muller2019crowdbreaks}.
Between July 1st, 2017 and October 1st, 2020 a total of \num{57.5}M tweets (including \num{39.7}M retweets) in English language by \num{9.9}M unique users were collected using the public filter stream endpoint of the Twitter API.
The tweets matched one or more of the keywords ``vaccine'', ``vaccination'', ``vaxxer'', ``vaxxed'', ``vaccinated'', ``vaccinating'', ``vacine'', ``overvaccinate'', ``undervaccinate'', ``unvaccinated''.
The data can be considered complete with respect to these keywords.

\subsection{Annotation data}
\label{sec:annotation_data}
Human annotation of a subset of tweets was performed through the Crowdbreaks platform~\parencite{muller2019crowdbreaks}.
Tweets were anonymized by replacing user mentions and URLs with placeholders.
Tweets between February 2nd 2018 and November 11th 2020 were sampled for annotation if they contained at least 3 words.
Exact duplicates were removed.
Annotators were asked the question ``What is the attitude of the author of the tweet regarding vaccines?'' and given the three options ``negative'',  ``neutral'', and ``positive''.
Annotation was performed both on Amazon Turk (mTurk) and, to a smaller extent (roughly 1\% of all annotations) by public users on the Crowdbreaks website.
We yield a dataset of \num{44843} annotations (Fleiss' kappa of \num{0.30}), which resulted in \num{11893} three-fold annotated tweets.
Tweets with less than two-third agreement were excluded and conflicts were decided through majority vote.

\subsection{Training of classifiers}
\label{sec:training}
In this work we leverage two different classifiers: FastText~\parencite{joulin2016bag} and BERT~\parencite{devlin2018bert}.
For both models, hyperparameters were first tuned on the full annotation data to yield optimal performance and then fixed for further experiments.
For FastText we used 10 dimensions, 500 epochs, a learning rate of \num{0.01}, and using 1-gram embeddings.
Optimal results were yielded by lower casing texts, converting them to ASCII and using the tags ``user'' and ``url'' for anonymization.
BERT models of the type bert-large-uncased (pretrained in English language) were trained for \num{20} epochs, training batch size of \num{32}, and a learning rate \num{2e-5} (using 10\% warmup with linear decay to zero), as recommended in recent literature~\parencite{mosbach2020stability,dodge2020fine}.
FastText models were trained on a university cluster using the Crowdbreaks \textsc{text-classification} library\footnote{\url{https://github.com/crowdbreaks/text-classification}} and BERT models were trained using Google Cloud v3-8 TPUs and the \textsc{COVID-Twitter-BERT} library\footnote{\url{https://github.com/digitalepidemiologylab/covid-twitter-bert}}~\parencite{muller2020covid}.
For the purpose of predictions, text was preprocessed using the respective preprocessing approach.

\paragraph{Data availability.}
All data and code can be found on our public GitHub repository \url{https://github.com/digitalepidemiologylab/concept_drift_paper}.

\paragraph{Author contributions.}
M.M.\ collected the data, designed the experiments and analysed the data.
M.M.\ and M.S.\ conceptualized the work and wrote the manuscript.

\paragraph{Acknowledgments.}
The authors would like to acknowledge Dr.\ Per Egil Kummervold and Dr.\ Burcu Tepekule for their valuable comments and discussions.

\paragraph{Competing interests.}
The authors declare no competing interests.

\paragraph{Funding.}
This work received funding through the Versatile Emerging infectious disease Observatory (VEO) grant as a part of the European Commission’s Horizon 2020 framework programme (grant agreement ID: 874735).
Compute resources (Cloud TPUs) were provided through Google’s TensorFlow Research Cloud and the work was supported through Google Cloud credits in the context of COVID-19-related research.

\printbibliography
\end{multicols}

%%%%%%%%%%%%%%%% SUPPLEMENTARY

\pagebreak
\beginsupplement

\begin{center}
  \noindent\rule{\textwidth}{1.5pt} \\
  \vspace{.2cm}
  \textsc{\Huge{Supplementary Material}} \\
  \vspace{.1cm}
  \noindent\rule{\textwidth}{1.5pt}\\
  \vspace{.5cm}
  \textsc{\large{Addressing machine learning concept drift reveals declining vaccine sentiment during the COVID-19 pandemic}}\\
  \vspace{.5cm}
  Martin M\"uller$^{1}$, Marcel Salath\'e$^{1}$\\
  {\itshape ${}^1$ Digital Epidemiology Lab, Ecole polytechnique fédérale de Lausanne (EPFL), 1202 Geneva, Switzerland\\}
\end{center}

\section{Supplementary figures}

\pagebreak

\pagebreak

\pagebreak

\end{document}